          [1999/12/02 1.01 (PWD)]
\documentclass[letterpaper,twocolumn]{esapub} 
\def \sss  {\scriptscriptstyle}
\def \bpc  {B_{\rm pc}}
\def \epm  {e^\pm}
\def \epar {E_\parallel}
\def \rhot {R_{\rm hot}}
\def \thot {T_{\rm hot}}
\def \icsbr{{\cal \scriptstyle E}^{\rm \sss ICS}_{\rm br}}
\def \ect  {\varepsilon_{\rm ct}}
\def \eb   {\varepsilon_{\sss B}}
\def \rns  {R_{\rm ns}}
\def\la{\hbox{\hspace{1.5mm}}\raise2pt
       \vbox{\hbox{$<$}}\lower2pt
       \vbox{\moveleft9.0pt\hbox{$\sim$ }}\hbox{\hskip 0.05mm}}

\usepackage{natbib}
\usepackage{graphicx}

\title{Model Spectra of Rotation Powered Pulsars in the INTEGRAL Range}
\author[1]{J. Dyks}
\author[1,2]{B. Rudak}
\affil[1]{Nicolaus Copernicus Astronomical Center, Toru\'n, Poland}
\affil[2]{TCfA of Nicolaus Copernicus University, Toru\'n, Poland}
\author[3]{T. Bulik}
\affil[3]{Nicolaus Copernicus Astronomical Center, Warsaw, Poland}

\begin{document}


\maketitle

\begin{abstract}
The energy range of IBIS is a promising ground for testing
mutual relations of distinct components expected in the spectra
of high-energy radiation from rotation powered pulsars. 
According to some polar-cap models
two such components - due to curvature and synchrotron emission - may contribute 
comparable amounts of power between $15$~keV and $10$~MeV
(Rudak \& Dyks, 1999). Zhang \& Harding (2000) argued recently for the inclusion
of a third possible component,
due to inverse Compton scattering (ICS) of soft thermal photons 
on secondary $\epm$-pairs. 
Here  
we present the results of Monte Carlo calculations of all three spectral components
within a polar-cap model which allows for interactions of relativistic particles
with the soft photons coming from the pulsar surface.
For teragauss pulsars with the surface temperature of a few$\times
10^5$~K the ICS component dominates the spectrum in the energy range
below $10$~MeV, and thus its presence increases the ratio of X-ray
to $\gamma$-ray luminosity (in comparison to the models 
ignoring the ICS on secondary $\epm$-pairs)
to a level observed in the Vela pulsar.
\end{abstract}

\section{Introduction}

The standard polar cap model of pulsars refers to the pair-photon
cascades induced by photons of curvature radiation (CR) 
due to ultrarelativistic
electrons leaving the neutron star (NS) surface (primary particles).
Although successful in reproducing gamma-ray spectra observed above $100$~MeV
\citep{dh82}
 the model fails to reproduce the relative level of X-ray and
gamma-ray emission. As shown by \citet{zh00} this might be
caused by
neglecting the resonant inverse Compton scattering (ICS) of thermal photons
from the surface by the secondary $\epm$ pairs.
The authors developed an approximate analytical method to 
model the X-ray luminosities (with the ICS contribution included)
of pulsars detected by ROSAT and ASCA.
Their results indicate that
in terms of power in the X-ray range the ICS component may be
competitive to both the curvature and synchrotron components. However,
in order to verify the conclusion of \citet{zh00}, and to obtain detailed spectra
the Monte Carlo simulations are necessary. 

In this work we present basic features of such Monte Carlo spectra
for different pulsar parameters. 
In particular, the role and properties of 
the spectral component due to resonant inverse
Compton scattering on secondary $\epm$-pairs are addressed. We find that the shape and the strength
of this component are sensitive to 
the strength of the local magnetic field as well as to the size and temperature of 
the thermal source.

\section{The model}

As a starting point we took a model of high-energy processes in pulsar magnetospheres
as proposed by \citet{dh82}.
The model was then furnished with a modern version
of a strong electric field $\epar$ after \citet{hm98}
with the lower, and the upper boundary of the accelerator at the height $h_0$, 
and $h_c$ above the stellar surface, respectively.
In each case the size of acceleration 
zone $h_c - h_0$ was determined
selfconsistently in the course of calculations. An extended source of soft thermal photons
located on the pulsar surface has been added (see below for details).
Primary electrons slide
along open dipolar lines of the magnetic field and
attain ultrarelativistic energies in the presence of this field. At the same time 
they loose the energy via two processes: curvature radiation (CR) and inverse
Compton scattering (ICS).
Due to a large strength of the $\epar$, the former process dominates the cooling
of the electrons. In consequence, vast majority of electromagnetic cascades 
is induced by curvature photons and very few by ICS (upscattered soft) photons. 
This allows us to ignore any ICS-induced
cascades and to model the cascades in the following way:

First,
the CR-induced cascade of $\epm$ pairs and synchrotron
photons is simulated following \citet{rd99}.
The primary electrons are accelerated along the last open 
lines of purely dipolar magnetic field and emit curvature photons. 
Some of these photons undergo one-photon magnetic absorption
producing $\epm$ pairs. The pairs emit in turn synchrotron photons (SR) which also 
may produce
higher generations of $\epm$ pairs with their own SR photons, etc..
  
In the second step, the simulation of the resonant ICS for the $\epm$ pairs
proceeds after \citet{dr00}.
Three different cases for the source of thermal photons are
considered:
{\bf A)} a hot cap of temperature $T_{\rm hot} \simeq {\rm a\ few} \times
10^6$~K
and radius $\rhot \simeq 10^5$ cm
centered on the magnetic pole;
{\bf B)} entire neutron star surface with $T_{\rm surf} \simeq {\rm a\ few}\times
10^5$;
{\bf C)} the hot cap of {\bf A)} with the
rest of the surface at $T=5\cdot 10^5$~K.
In order to simplify the calculations of scatterings in the cases
A) and C) (with the hot cap) we made the following axial geometry approximation:
for the local ambient soft photon
field as seen by a $\epm$-pair we took a field that would be seen 
if the pair were moving along the dipole axis
at the same radial coordinate. 

For the sake of completeness the ICS component due to the primary electrons is
calculated
in the last step, even though this component is not important
energetically.

We would like to emphasize that the numerical code
constructed for this model does not
follow the directional pattern of the outgoing radiation.
It calculates a `single particle' spectrum of a pulsar, ie.~the spectrum
due to a single primary electron accelerated to an ultrarelativistic energy 
in the presence of a strong electric field.
In this sense our work should be considered as a first approximation 
in modelling phase-averaged spectra.

\section{Results}

\begin{figure}
\includegraphics[width=1.0\linewidth]{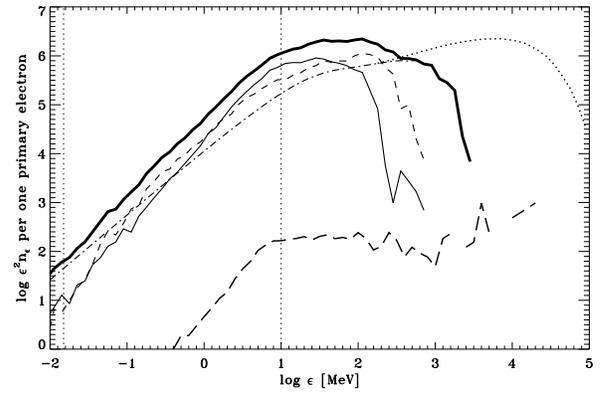}
\caption{The radiation energy spectrum per logarithmic energy bandwidth
due to a single primary electron. The pulsar parameters are
$\bpc = 6\cdot 10^{12}$~G, and $P=0.1$~s and the case {\bf A} ($\rhot=10^5$
cm, $\thot = 5\cdot 10^6$~K) is considered.
The spectrum consists of four components: the curvature radiation
(dot-dashed)
and the ICS (long dashed) are due to the primary electron, while 
the synchrotron radiation (short dashed)  and the ICS
(thin solid) are due to the secondary $\epm$ pairs.
Thick solid line indicates superposition of the CR, the SR and the $\epm$
ICS components.
The absorbed part of the curvature spectrum
is marked with dotted line.
The dotted verticals mark the energy range of IBIS.
}
\label{fig1}
\end{figure}

Fig.1 shows the spectrum calculated   
for a pulsar with 
$\bpc = 6$~TG, $P = 0.1$~s
(the parameters close to those of the Vela or B1706$-$44)
and a hot cap with 
$R_{\rm hot} = 10^5$ cm and $T_{\rm hot} = 5 \cdot 10^6$~K.
The total spectrum
(thick solid line) is a sum of three components: the curvature spectrum
(dot-dashed line), 
the synchrotron spectrum emitted by the $e^\pm$ pairs (short dashed line),
and the inverse Compton spectrum of photons scattered by the $e^\pm$ pairs
(thin solid line).
The fourth spectral component --
the inverse Compton spectrum of photons upscattered by the
primary electron -- shown as the long dashed line is never energeticaly important.
This is because the primary undergoes a strong acceleration
and its energy increases quickly above the value required to fulfill the
resonant condition. 
For this reason we have also ignored any magnetic absorption of these ICS photons 
and the subsequent creation of $\epm$-pairs.
At the same time
the ICS component due to the $\epm$ pairs is comparable in power to the CR and SR
components.
The secondary $e^\pm$ pairs do not accelerate
and remain in resonance until they lose most of their energy
via resonant scatterings.

The $e^\pm$-ICS component 
(thin solid line in Fig.~\ref{fig1}) has a well-known shape worked out 
analytically by \citet{dermer90}
for the `thick target' regime of the ICS ie.~for the case when scatterings
considerably decrease the energy of the scattering particle. It resembles
a broken power law with  
the break located at $\icsbr \simeq B_{12}^2/ T_6$~MeV, where $B_{12} =
B/(10^{12}\ {\rm G})$ and $T_6 = T/(10^6\ {\rm K})$.
Below the break
the photon index $\alpha_{\rm \sss ICS}$ assumes the value close to $0$, while above the break 
$\alpha_{\rm \sss ICS} 
\simeq -2$  (it is the same as for
the energy distribution of the ICS emitting pairs, cf.
\citeauthor{dh89}, \citeyear{dh89}; \citeauthor{dermer90}, \citeyear{dermer90}).
The shape of the synchrotron component is somewhat more
complicated (see \citet{rd99} for detailed analysis), 
nevertheless its level lowers considerably 
below
the blueshifted cyclotron energy $\ect \simeq \eb\gamma_\parallel \sim 3\ 
{\rm MeV} (P/(0.1\ {\rm s}))^{1/2} (\bpc/{\rm TG})$
(\citeauthor{rd99}, \citeyear{rd99}; \citeauthor{hd99}, \citeyear{hd99}).
With the CR break at $\sim 10^2$~MeV \citep{rd99}, most of the power contained
in all these spectral components in the case shown in Fig.~\ref{fig1} 
is confined to the energy range above $\sim 10$~MeV.

This paucity of  power in hard X-ray/soft gamma-ray range, clearly visible in Fig.1,
stays in disagreement with observed spectra
of high-$B$ (ie.~classical) pulsars
(note the extreme case of the Crab pulsar with a comparable energy output
within soft X-ray and hard gamma-ray window, \citeauthor{thompson97},
\citeyear{thompson97}).

\begin{figure}[t]
\includegraphics[width=\linewidth]{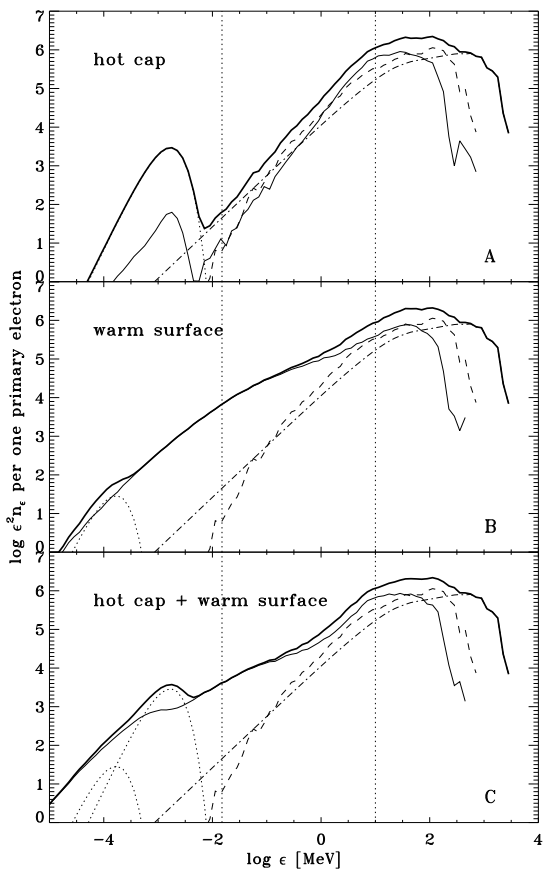}
\caption{The radiation energy spectra per logarithmic energy bandwidth
for three cases of the thermal photon source 
mentioned in the text:
{\bf A} - the hot cap with
$\rhot = 10^5$ cm and $\thot = 5\cdot 10^6$~K;
{\bf B} - the entire surface at $T_{\rm surf}=5\cdot 10^5$~K;
{\bf C} - the hot cap of {\bf A} with the
rest of the surface at $T=5\cdot 10^5$~K.
In all cases $\bpc = 6\cdot 10^{12}$~G and $P=0.1$~s.
The dot-dashed line is for the curvature component, the dashed line is
the synchrotron component and the thin solid line denotes the inverse
Compton spectrum of photons scattered by the $\epm$ pairs.
Dotted line presents contribution of thermal radiation from the surface.
Thick solid line is the total spectrum
per single primary electron. 
}
\label{fig2}
\end{figure}

\begin{figure}[t]
\includegraphics[width=\linewidth]{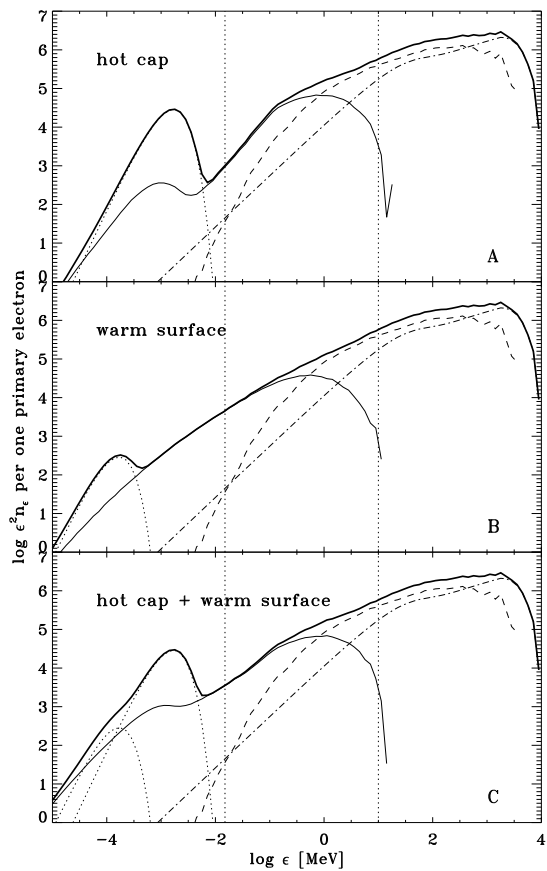}
\caption{The radiation energy spectra per logarithmic energy bandwidth
for the same cases of the thermal photon source
as in Fig.~\ref{fig2} but for $\bpc = 0.6\cdot10^{12}$~G and $P=0.1$~s.
The meaning of all lines is the same as in Fig.~\ref{fig2}.
}
\label{fig3}
\end{figure}

In order to increase the spectrum level in this range we will proceed
with three different modifications of the case presented in Fig.1.

First, we will change the size and temperature of the soft-photon source, 
ie.~we
will consider the cases {\bf B} and {\bf C}.
In Fig.~2  the spectrum of Fig.~1 
is presented with more details (note the wider energy range) in the upper panel and 
compared with two spectra calculated for case {\bf B} 
(the middle panel) and case {\bf C} (the lower panel).
The dotted lines in Fig.~2 present the blackbody spectrum of the thermal
photon source. Its normalization was estimated in an approximate way
described elsewhere (Rudak \& Dyks (2001), in preparation). 
The other lines have the same meaning as in Fig.~1.
Note that the ICS component for the warm surface case
(thin solid line in the middle panel) extends well
into the X-ray regime and at $10$~keV
its level exceeds by two orders of magnitude a level of the spectrum
for the hot cap case (upper panel). 
It has no longer a single power-law shape.
The significant difference between the shape of the ICS component for the
hot cap case (upper panel in Fig.~2) and the warm surface case (middle
panel) comes from a difference between the energy loss rates of electrons
in these two cases.
The energy-loss-scale-length  $\lambda_{\dot\gamma}$
is at the resonance of the order of $10^2$ cm and $10^4$ cm in the two cases, 
respectively.
In the hot cap case ({\bf A}) electrons lose most energy over a very short distance
from the creation
point and then leave the resonance regime.
In the warm surface case the pairs lose energy in a much slower rate and fulfill the
resonant condition
over the substantially longer length scale.
Because of the larger size of the
thermal photon source (the entire surface)
they can still scatter resonantly even at a few stellar radii above the
surface
where the soft tail of the ICS component is produced.
When both the hot cap and the rest of surface emit thermal X-ray photons,
then both the break at $\icsbr$ and the soft tail are present in the ICS
component of the spectrum
(the case shown in the lower panel of Fig.~\ref{fig2}).

Second, we can force the spectrum to extend towards lower energies 
by either lowering the strength of $\bpc$ at the polar cap or elevating
the accelerator  to some high altitude ($h_0 > 0$) to lower the local magnetic field
strength.
The latter case  is justified by the idea
of the high altitude accelerator as introduced and argued for by
\citet{hm98}. 
It is the $e^\pm$-ICS component which reacts in this desirable way when the
strength of $B$ decreases:
this component extends towards lower photon energies faster than the SR component
($\icsbr \propto B^2/T$ whereas $\ect \propto B$).

\begin{figure}[t]
\includegraphics[width=\linewidth]{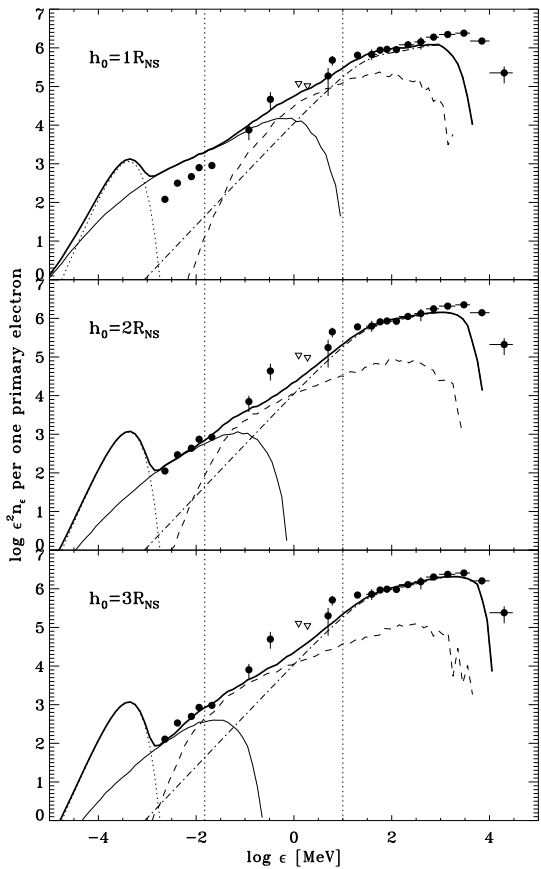}
\caption{The radiation energy spectra of the Vela pulsar
($\bpc = 6\cdot 10^{12}$~G, $P=89$~ms, $T_{\rm surf} = 1.26\cdot 10^6$~K).
Three panels correspond to different altitudes of a lower boundary
of acceleration zone: $h_0=1$, $2$, and $3\ \rns$ above the surface.
The meaning of all lines is the same as in Fig.~\ref{fig2}.
Filled dots (some with error bars)
present data from RXTE, OSSE, COMPTEL and EGRET
(\citet{thompson97},
\citet{shd99}). Triangles denote OSSE and
COMPTEL upper limits.
}
\label{fig4}
\end{figure}

Let us lower the magnetic field strength at the polar cap by one order
of magnitude: $\bpc  = 0.6\cdot 10^{12}$~G. The spectra for 
the cases {\bf A}, {\bf B} and {\bf C} are now presented in Fig.~\ref{fig3}.
In accord with the relation $\icsbr \propto B^2/T$, 
the ICS component is shifted towards the lower
photon energy by two orders of magnitude and dominates for ${\cal
\scriptstyle E} \la 1$~MeV.
Again, the break at $\icsbr$ is present for the hot cap case ($T =
5\cdot 10^6$~K, upper panel) and
there is a soft tail and no sign of the break for the warm surface case
($T = 5\cdot 10^5$~K).
The  level of the $\epm$-ICS component for $\bpc = 0.6\cdot 10^{12}$~G is one order of
magnitude lower than the corresponding level for $\bpc = 6\cdot 10^{12}$~G.
We emphasize that the main reason for this effect
is not a low value of the energy loss rate
due to the ICS 
in weak magnetic fields 
(in {\em both} cases the pairs lose almost 100\% of energy due to
scatterings) but
the smaller total energy
content of the $\gamma_\parallel mc^2$ distribution for cascade pairs in the
low-$B$ case \citep[cf.][]{zh00}.
This is also the main reason for which the $\epm$-ICS component is
negligible for millisecond pulsars (ie.~the photon index predicted for the
INTEGRAL range is that of synchrotron component: $\alpha_{\rm \sss SR} =
-1.5$, \citet{rd99}).

Fig.~\ref{fig4} is to illustrate the case of elevating
the accelerator  to some high altitude. Three subcases, with
$h_0 =  1$, $2$, and $3\ \rns$ were considered.
Moreover, the figure enables the comparison of the model predictions 
with the data for the Vela pulsar, coming  
from the compilation by \citet{thompson97} and \citet{shd99}.
The spectra were calculated for the entire surface
at $T_{\rm surf} = 1.26
\cdot 10^6$~K \citep{ogelman95}. In spite of large
altitudes at which the $\epm$ pairs are created (low $B$),
the $\epm$ ICS component
contributes considerably within the X-ray range
and improves agreement between the theory and the data.

\section*{Acknowledgments}
This work was supported by KBN grant 2P03D 02117.


\end{document}